\documentclass[twocolumn,preprintnumbers,amsmath,amssymb,superscriptaddress,10pt]{revtex4}
\usepackage{amssymb}
\usepackage{graphicx,color}
\usepackage{bm}
\usepackage{ulem}

\begin{document}
\title{Shear viscosity of ultrarelativistic Boson systems in the presence \\ of a Bose-Einstein condensate}

\author{Zhengyu Chen}
\affiliation{Department of Physics, Tsinghua University and Collaborative Innovation Center of Quantum Matter, Beijing 100084, China}

\author{Carsten Greiner}
\affiliation{Institut f$\ddot{u}$r Theoretische Physik, Johann Wolfgang Goethe-Universit$\ddot{a}$t Frankfurt, Max-von-Laue-Strasse 1, 60438 Frankfurt am Main, Germany}

\author{Zhe Xu \footnote{xuzhe@mail.tsinghua.edu.cn}}
\affiliation{Department of Physics, Tsinghua University and Collaborative Innovation Center of Quantum Matter, Beijing 100084, China}

\author{Pengfei Zhuang}
\affiliation{Department of Physics, Tsinghua University and Collaborative Innovation Center of Quantum Matter, Beijing 100084, China}

\begin{abstract}
We calculate for the first time the shear viscosity of ultrarelativistic Boson systems in the presence
of a Bose-Einstein condensate (BEC).
Two different methods are used. One is the Grad's method of moments and another is the
Green-Kubo relation within a kinetic transport approach. In this work we consider 
a Boson system with isotropic elastic collisions and a gluon system with elastic scatterings
described by perturbation QCD (pQCD). The results show that the presence
of a BEC lowers the shear viscosity. This effect becomes stronger for the increasing proportion
of the BEC in the Boson system and is insensitive to the detail of interactions.
\end{abstract}


\maketitle

\section{Introduction}
\label{sec1:intro}
The experiments of heavy-ion collisions at the Relativistic Heavy Ion Collider (RHIC) and
at the Large Hadron Collider (LHC) \cite{Adams:2005dq,Adcox:2004mh,Aamodt:2010pa} showed
strong indications for the formation of a new thermal state of matter composed of quarks
and gluons, the quark-gluon plasma (QGP). Theoretically, the color glass condensate (CGC)
effective field theory \cite{McLerran:1993ni} can describe the evolution of the initially
freed gluons to a glasma \cite{Gelis:2010nm,Lappi:2006fp,Weigert:2005us}, which is still
far from thermal equilibrium. The formation of QGP from this nonequilibrium glasma is
a dynamical thermalization process, which is not understood yet from the first principles
within QCD. Efforts from recent investigations \cite{Blaizot:2011xf,Blaizot:2012qd} showed
that gluons in the glasma can be highly overpopulated. As a fundamental consequence of
quantum statistics, a gluon BEC could appear. The gluon condensation, if it occurs,
will accelerate the thermalization process and thus, is regarded as a promising mechanism
for the fast thermalization of quarks and gluons produced in heavy-ion collisions at RHIC
and LHC. Many recent studies have been devoted to the nonequilibrium dynamics and BEC
formation within either kinetic approach 
\cite{Blaizot:2013lga,Blaizot:2014sha,Blaizot:2014jna,Huang:2013lia,Scardina:2014gxa,Xu:2014ega,Blaizot:2015wga,Blaizot:2015xga,Meistrenko:2015mda,Zhou:2017zql}
or classical field theory \cite{Berges:2013fga,Berges:2013eia,Kurkela:2012hp,Schenke:2016ksl}.

Whether or not gluon BEC can be formed in heavy-ion collisions, is still under debate
\cite{Huang:2014iwa,Huang:2013lia,Blaizot:2016iir,Kurkela:2014tea,Kurkela:2011ti}. 
In this work we will not touch this issue. We assume the existence of a gluon BEC and
would like to study its effect on the shear viscosity of gluons.

The shear viscosity is an important quantity manifesting the transport property of
the QGP. Extracted from the flow measurements at RHIC and LHC by comparing with
the viscous hydrodynamic calculations, one found small numbers of the shear viscosity
over the entropy density ratio ($\eta/s$) of the QGP \cite{Schenke:2011tv}.
To understand the nature of the small $\eta/s$, is one of the motivations for this work. 

From kinetic theory we can realize that stronger interactions will lead to faster
thermalization as well as smaller shear viscosity \cite{Xu:2007ns}. When the gluon
condensation accelerates thermalization, it is expected that processes corresponding
to the gluon condensation will lower the shear viscosity. However, quantitatively it is
nontrivial to confirm this expectation, especially for massless Boson systems.
In Sec. \ref{sec2:bec} we discuss the description of BEC in kinetic theory from binary
processes, especially highlight the rate of condensation. The collision rate of processes
involving the BEC is infinitely large, while the collision angle is zero. How these extreme
processes affect the shear viscosity is one key ingredient of this work.

In Sec. \ref{sec3:grad} we derive for the first time the shear viscosity of massless
bosons in the presence of a BEC analytically by applying the Grad's method of
moments \cite{Grad1949,deGroot,Muronga:2003ta,Muronga:2006zx}.
As a first test case, we assume elastic collisions with constant cross sections and
isotropic distribution of collision angles. This is customary in kinetic theory calculations.
Later in Sec. \ref{sec5:gluon}  we will relax this restriction. As a complementary method,
we numerically calculate the shear viscosity in Sec. \ref{sec4:greenkubo} for the same
system and with the same interactions by using the Green-Kubo relations within the
Boltzmann Approach of MultiParton Scatterings (BAMPS) \cite{Wesp:2011yy}.
The two independent methods agree perfectly and confirm our
analytic and numerical calculations. In Sec. \ref{sec5:gluon}  we consider a realistic
gluon system, employing binary pQCD cross sections in the presence of a gluon
BEC within the numerical framework BAMPS. We will summarize in Sec. \ref{sec6:sum}.
The details of the thermodynamic integrals needed in Sec. \ref{sec3:grad}
are given in Appendix.

\section{Kinetic description of Bose-Einstein condensation}
\label{sec2:bec}
In this section we give a brief description of Bose-Einstein condensation by
using the kinetic Boltzmann equation. The detailed description can be found
in Ref. \cite{Zhou:2017zql}. 

The one-particle phase space distribution function
$f(x,p)$ is decomposed into two parts $f=f^g+f^c$, where $f^g$ denotes the
distribution of gas (noncondensate) particles and
$f^c=(2\pi)^3 n_c \delta^{(3)}({\mathbf p})$ denotes the distribution of the
condensate particles with zero momentum. $n_c(x)$ is the local particle density
of the condensate. In this work we consider elastic collisions only. 
This assumption is reasonable for systems with a conserved particle number.
For a gluon system, however, number-changing processes may have to be taken
into account. In Sec. \ref{sec5:gluon} we will give a short discussion about why
we do not consider number-changing processes of gluons in the present work.

Denoting gas particles by $g$ and condensate particles by $c$, we consider
$g+g \to g+g$, $g+c \to g+g$, and $g+g \to g+c$ processes. The Boltzmann
equations for gas and condensate particles are then given as follows:
\begin{eqnarray}
\label{be1}
&&p_1^{\mu} \partial_{\mu} f^g_1(x,p_1)=C[f_i(x,p_i)]\nonumber \\
&& = \frac{1}{2} \int d\Gamma_2 \frac{1}{2} \int d\Gamma_3 d\Gamma_4
| {\cal M}_{34\to 12} |^2 \nonumber \\
&& \times \ \left[ f^g_3 f^g_4 (1+f^g_1) (1+f^g_2) 
+f^g_3 f^g_4 (1+f^g_1) f^c_2 \right. \nonumber \\
&& \ \ \ \ + f^c_3 f^g_4 (1+f^g_1) (1+f^g_2)
+f^g_3 f^c_4 (1+f^g_1) (1+f^g_2) \nonumber \\
&& \ \ \ \  - f^g_1 f^g_2 (1+f^g_3) (1+f^g_4) - f^g_1 f^c_2 (1+f^g_3) (1+f^g_4)
\nonumber \\
&& \ \ \ \, \left. - f^g_1 f^g_2 f^c_3 (1+f^g_4) - f^g_1 f^g_2 (1+f^g_3) f^c_4 \right] 
\nonumber \\
&& \times (2\pi)^4 \delta^{(4)} (p_3+p_4-p_1-p_2) \,,
\end{eqnarray}
\begin{eqnarray}
\label{be2}
&&p_1^{\mu} \partial_{\mu} f^c_1(x,p_1)=\frac{1}{2} 
\int d\Gamma_2 \frac{1}{2} \int d\Gamma_3 d\Gamma_4
| {\cal M}_{34\to 12} |^2 \nonumber \\
&& \times \ \left[ f^g_3 f^g_4 f^c_1 (1+f^g_2) 
- f^c_1 f^g_2 (1+f^g_3) (1+f^g_4) \right] \nonumber \\
&& \times (2\pi)^4 \delta^{(4)} (p_3+p_4-p_1-p_2) \,,
\end{eqnarray}
where $d\Gamma_i=d^3p_i/(2E_i)/(2\pi)^3$ and $f_i^{g/c}=f_i^{g/c}(x,p_i)$ with 
$i=1,2,3,4$ is the distribution function of $i$th particle. 
Elastic collisions $34\to12$ and $12\to34$ are determined by the collision kernel
$| {\cal M}_{34\to 12} |^2$ and  $| {\cal M}_{12\to 34} |^2$, which are equal.

The rate of the Bose-Einstein condensation can be obtained by the integration
of the right-hand side of Eq. (\ref{be2}) over $d\Gamma_1$ in the local rest frame.
The details of the integration can be found in Ref. \cite{Zhou:2017zql}. The final
result is
\begin{eqnarray}
\label{condrate}
\frac{\partial n_c}{\partial t} = \frac{n_c}{64\pi^3} 
&\int& dE_3 dE_4 \left [ f_3^g f_4^g - f_2^g (1+f_3^g+f_4^g) \right ] 
\nonumber \\
&& \times \ E \left[ \frac{| {\cal M}_{34\to 12} |^2}{s} \right]_{s=2mE} \,.
\end{eqnarray}
The two terms on the right-hand side of Eq. (\ref{condrate}) correspond to
kinetic processes for the condensation and the evaporation, respectively.
$E=E_3+E_4$ is the total energy in the collision, 
$P=|{\mathbf p}_3+{\mathbf p}_4|$ is the total momentum, and $s=E^2-P^2$ is
the invariant mass. $m$ denotes the particle mass at rest, which is set to zero
throughout this paper. We found \cite{Zhou:2017zql} that in order to describe
the condensation of massless bosons with a finite rate, the ratio 
$|{\cal M}_{34\to 12} |^2/s$ at $s=0$ should be nonzero and finite.
For isotropic collisions, i.e., the distribution of the collision angle is
isotropic, we get $| {\cal M}_{34\to 12} |^2= 32 \pi s \sigma_{22}$, where 
$\sigma_{22}$ is the total cross section. We see that isotropic collisions
with finite cross sections can describe the condensation process of massless
bosons with a finite rate.

At thermal equilibrium, $n_c$ does not change with time. The two integrals
in the right-hand side of Eq. (\ref{condrate}) cancel. More important is that
both integrals will be infinitely large, when $f^g_i$ takes the fully equilibrated
Bose-Einstein distribution function, $f^g_i=1/(e^{E_i/T}-1)$, where $T$ is
the temperature, because for instance, $\int_m^{\infty} dE_3 f_3^g$ is
logarithmically divergent for $m=0$ (which also holds for finite mass).
This indicates that both the rates of $g +g \to g +c $ and $g+c \to g+g$ are
infinitely large at equilibrium. Since the main purpose of this work is to
calculate the shear viscosity of a Bose gas in the presence of a BEC, 
one may argue that the infinite collision rates of $g +g \to g +c $
and $g+c \to g+g$ would naively lead to a vanishing shear viscosity of the
gas (noncondensate) particles. However, the collision rate is not the only
determinant of the magnitude of the shear viscosity. Another determinant is
the collision angle. In processes $g +g \to g +c $ and $g+c \to g+g$ with
massless particles, all the momenta of the noncondensate particles before
and after the collision should be parallel. The only change after
$g +g \to g +c $ and $g+c \to g+g$ processes are the magnitudes of the
momenta, but not their directions. This corresponds to zero collision angle.
One may argue again that those collisions would not contribute to the shear
viscosity. We see that the infinite collision rate and zero collision angle are
two extremes in the condensation processes. How the counterbalance of
these two extremes will affect the shear viscosity is an interesting and
nontrivial issue, which will be addressed in the following sections with two different
methods. One is analytical from second-order kinetic theory \cite{Muronga:2006zx},
while another is numerical from the Green-Kubo relation within the transport
approach BAMPS \cite{Wesp:2011yy}.

\section{Shear viscosity coefficient from second-order kinetic theory}
\label{sec3:grad}
Relativistic causal dissipative hydrodynamic equations can be derived from
the kinetic theory by applying Grad's method of  moments \cite{Grad1949}.
A detailed prescription for the derivation of shear viscosity from the
second-order kinetic theory is given in Ref. \cite{Muronga:2006zx}.  

In this work we use the general formula of the second-order shear viscosity
derived in Ref. \cite{Muronga:2006zx} for a special case: a massless Boson
system in one-dimensional Bjorken expansion \cite{Bjorken:1982qr}. The local
equilibrium distribution function of bosons is the Bose-Einstein distribution
function:
\begin{equation}
f_{BE}(x,p)= \frac{1}{e^{u_\mu p^\mu/T}-1}\,,
\label{fbe}
\end{equation}
where $u^\mu(x)$ is the four-fluid velocity. For a one-dimensional Bjorken
expansion, we have $u^\mu=(t, 0, 0, z)/\tau$ with $\tau=\sqrt{t^2-z^2}$
\cite{Bjorken:1982qr}. In the local rest frame, $u^\mu=(1,0,0,0)$.

The shear viscosity is a material property. Its value does not depend on
the special form of the collective motion of the matter. The assumption
of the one-dimensional Bjorken expansion will simplify the calculation
of the shear viscosity, because in this case the heat flux $q^\mu$ vanishes
in the local rest frame \cite{El:2008yy}. In addition, for massless systems,
the bulk pressure $\Pi$ becomes zero \cite{El:2008yy}. Then, the formula of
the second-order shear viscosity from Ref. \cite{Muronga:2006zx} is reduced to
\begin{equation}
\eta =-\frac{ \pi_{\alpha\beta}\pi^{\alpha\beta}}{2 T  C_0 \pi_{\mu\nu}P^{\mu\nu}}\,.
\label{eta}
\end{equation}
$\pi^{\alpha\beta}$ denotes the shear tensor. We express $C_0$ explicitly as given
in Ref. \cite{Muronga:2006zx}: 
\begin{equation}
C_0^{-1} = \frac{2}{15} \int \frac{d^3 p}{(2\pi)^3 p^0} \left [(u_\alpha p^\alpha)^2
-p_\gamma p^\gamma \right ]^2 \frac {e^{u_\nu p^\nu/T}}{({e^{u_\nu p^\nu/T}}-1)^2} \,.
\label{c0}
\end{equation}  
The integral can be easily performed and we obtain $C_0= \pi^2/(8\zeta[5] T^6)$,
where $\zeta[x]$ is the zeta-function. The functional $P^{\mu\nu}$ that is
\begin{equation}
P^{\mu\nu} =\int \frac{d^3 p_1}{(2\pi)^3 p^0_1} p^\mu_1 p^\nu_1 C[f_i(x,p_i)]
\label{Pmunu}
\end{equation}
involves interactions between particles by means of the collision term in
the Boltzmann equation (\ref{be1}). The viscosity is the measure of the medium
response to a small disturbance, which leads the system to deviate slightly
from local equilibrium. The one-particle phase-space distribution function
$f(x,p)$ has then the following form (the subscripts are omitted),
\begin{equation}
f(x,p)=f_{BE}(x,p)\left \{ 1+ \left [1+f_{BE}(x,p) \right ] \phi(x,p) \right \}\,.
\label{f_expansion}
\end{equation}
Using the relativistic Grad's 14-moment approximation \cite{Grad1949,IS} or
variational method \cite{deGroot}, $\phi(x.p)$ is approximated up to the
second order of momentum \cite{Muronga:2006zx,El:2008yy},
\begin{equation}
\phi(x,p) = C_0\pi_{\mu\nu}p^\mu p^\nu \,.
\label{eq:phieq}
\end{equation}
Since we have assumed one-dimensional Bjorken expansion, in the local rest frame,
the shear tensor takes the form 
$\pi^{\mu\nu}=diag(0, -{\bar \pi}/2, -{\bar \pi}/2, {\bar \pi})$.
According to Eqs. (\ref{Pmunu})-(\ref{eq:phieq}), 
$P^{\mu\nu}$ will be calculated when putting $\pi^{\mu\nu}$ into $\phi(x,p)$.
We will see in Appendix that $\pi_{\mu\nu}P^{\mu\nu}$ is proportional to
${\bar \pi}^2$ and thus, ${\bar \pi}$ from the numerator and denominator of
Eq. (\ref{eta}) cancels out. The shear viscosity does not depend on ${\bar \pi}$
and can be evaluated explicitly, if the matrix elements of particle interactions,
which are involved in the collision term $C[f_i(x,p_i)]$, are given.

We now calculate for the first time the second-order shear viscosity of the
noncondensate particles in the presence of a Bose-Einstein condensate.
We assume isotropic collisions with a constant cross section. A constant cross
section means that the total cross section is independent of $s$.
The assumption is useful for testing the numerical framework, and is always
interesting, because in kinetic theory constant isotropic cross sections are
very often used. With this assumption, some integrals in Eq. (\ref{Pmunu})
can be carried out analytically. The rest has to be computed numerically.
We obtain
\begin{equation}
\label{eta_secord}
\eta= k \frac{T}{\sigma_{22}} \,,
\end{equation}
where
\begin{eqnarray}
k= \frac{48\zeta[5]^2}{\pi^4} && \left [ \frac{8}{45}\zeta[5]-\frac{8}{63} \pi^2\zeta[3]
 +1.827 \right. \nonumber \\
&& \left. +\frac{4n_c}{5\pi^2 T^3} \left ( 12\zeta[3]^2 -1.504 \right ) \right ]^{-1}
\,.
\label{eq:viscoBEC}
\end{eqnarray}
The result, Eqs.  (\ref{eta_secord}) and (\ref{eq:viscoBEC}), is new and nontrivial.
The details of the integration are presented in Appendix.
 
The shear viscosity is proportional to the temperature and inversely
proportional to the total cross section. For the absence of the condensate
($n_c=0$) we obtain $k=1.05$. This value is smaller than that ($k=1.2$)
when assuming Boltzmann statistics (neglecting the Bose factors) \cite{El:2012cr}.
This shows that the Bose factors increase the collision rate, which
lowers the shear viscosity.

\begin{figure}[ht]
 \centering
 \includegraphics[width=0.45\textwidth]{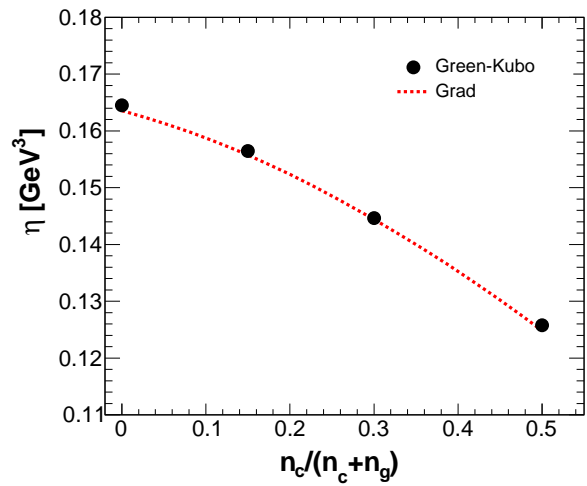}
\caption{The shear viscosity of a massless Boson system as
a function of $n_c/(n_c+n_g)$. The result from the second-order kinetic theory,
Eq. (\ref{eta_secord}), is shown by the red dashed curve, while the results
from the Green-Kubo formalism are depicted by the symbols.}
\label{fig:perc}      
\end{figure}

From Eqs.  (\ref{eta_secord}) and (\ref{eq:viscoBEC}) we realize that the
presence of the BEC lowers the shear viscosity. Since the number density of
noncondensate particles is $n_g=\zeta[3] T^3/\pi^2$, $\eta(n_c)/\eta(n_c=0)$
only depends on $n_c/n_g$. In other words, the larger the proportion of the
BEC in the system, the stronger is the effect of the BEC on the shear viscosity. 
By keeping $n_g$ constant (with a fixed temperature) and varying $n_c$, 
the proportion of the BEC in the system $n_c/(n_c+n_g)$ will change
accordingly. With $T= 0.4 \mbox{ GeV}$ and $\sigma_{22}=1 \mbox{ mb}$
we show in Fig. \ref{fig:perc} the shear viscosity as a function of
$n_c/(n_c+n_g)$ by the red dashed curve. We see a decreasing shear
viscosity when increasing $n_c$.

The main result of this work is that we find a finite and nonzero contribution
of the BEC (or $g+g\to g+c$ and $g+c\to g+g$ processes) to the shear
viscosity of massless noncondensate particles. Remember [see Eq. (\ref{condrate})]
that the rate for the condensation ($g+g\to g+c$) at equilibrium is infinite.
The logarithmic divergence comes from the integration
$\int_0^\infty dE f_{BE}$, since $f_{BE} \sim T/E$ at $E\to 0$.  Because
the same divergence appears in the rate for the evaporation ($g+c\to g+g$),
the net rate for the condensation at equilibrium is zero. This shows that
the gain and loss term cancel, when the collision term of the Boltzmann
equation is integrated in momentum space. In the case of the calculation
of the shear viscosity [see Eqs. (\ref{eta}) and (\ref{Pmunu})], the second
moment of the collision term instead of the collision term itself is integrated.
The gain and loss terms do not cancel, but lead to a net momentum transfer.
In the single gain and loss terms, there are still divergences due to the same
reason as in the calculation of the collision rates. We sum all gain and loss
terms with divergences and find that the term corresponding to the
momentum transfer in processes $1+2 \leftrightarrow 3+4$ is 
$E_1^2-(E_3^2+E_4^2)$. Since particle $2$ is the condensate particle
with $E_2=0$, we have $E_1=E_3+E_4$ due to the energy conservation.
Therefore, $E_1^2-(E_3^2+E_4^2)=2E_3E_4$ and there are only
integration with the mixed term and no integration such as 
$\int_0^\infty dE_{3,4} f_{BE}^{3,4}$. The divergences in the gain and loss
terms cancel with each other. The net momentum transfer is finite and nonzero.

Qualitatively,  collinear collisions involving a massless BEC particle ($g+g\to g+c$
and $g+c\to g+g$ processes) lead to a redistribution of the magnitude of
momentum, which is obviously a momentum degradation among different
nearby fluid layers, although the collision angle in such collinear collisions
is zero. The redistribution of the magnitude of momentum gives a nonzero
momentum transfer and a nonzero contribution to the entropy production. 
Therefore, collinear collisions involving a BEC particle should contribute
to the shear viscosity, as demonstrated mathematically in Appendix. 
In the next section we will use a different method to calculate the shear viscosity
in the presence of a BEC, in order to prove the result shown in this section.

\section{Shear viscosity coefficient from Green-Kubo relations}
\label{sec4:greenkubo}
According to the several works by Green and Kubo \cite{Green1954,Kubo:1957mj},
which are motivated by Onsager's regression hypothesis \cite{Onsager1931},
transport coefficients can be related to the correlation function of the
corresponding flux or tensor in thermal equilibrium.  The Green-Kubo relation
for the shear viscosity is
\begin{equation} 
\label{green_kubo_definition} 
\eta = \frac{1}{10 \text{T}} \int_{0}^{+ \infty} \mathrm{d}t \int_{V}^{} \mathrm{d}^3 r \ 
\langle \pi^{ij}({\bf r}, t) \pi^{ij}(0,0) \rangle \,,
\end{equation}
where $i,j=x,y,z$ and $\langle \cdots \rangle$ denotes the ensemble average
in thermal equilibrium. The sum over $i$ and $j$ gives 
$\langle \pi^{ij}({\bf r}, t) \pi^{ij}(0,0) \rangle=10 \langle \pi^{xy}({\bf r}, t) \pi^{xy}(0,0) \rangle$.

Different from the one-dimensional Bjorken expansion assumed in the previous
section, we consider here a homogeneous static system in global thermal
equilibrium. At thermal equilibrium, fluctuations are still present. Thus,
$\pi^{xy}({\bf r},t)$ fluctuates around zero. The dissipation of fluctuations
leads to the relaxation of the correlation 
$\langle \pi^{xy}({\bf r}, t) \pi^{xy}(0,0) \rangle$ that is determined by
the shear viscosity as indicated in Eq. (\ref{green_kubo_definition}).

In this work, fluctuations are realized in BAMPS \cite{Xu:2004mz}, where
test particles are used to represent the particle phase space distribution
function. Although the number of test particles should be high enough to
ensure the high accuracy of the solution, it is still finite, which then
leads to fluctuations of $\pi^{xy}({\bf r},t)$. A calculation of the shear
viscosity using Green-Kubo relations within BAMPS (but without Bose statistics
and BEC) has been done in Ref. \cite{Wesp:2011yy}. We follow this framework,
but employ the newly developed BAMPS including Bose statistics and BEC
\cite{Zhou:2017zql}. The most details on the numerical implementations can be
found in Refs. \cite{Zhou:2017zql,Wesp:2011yy,Greif:2014oia}. In the following
we briefly present the numerical procedure and show some new numerical
implementations.

We consider a static box with volume $V=L^3$, where $L$ is the side length. 
The densities of test particles in the box are the physical particle number densities
($n_g$ and $n_c$) multiplied by $N_{test}$, which then indicates the number of
test particles per real particle. The condensate test particles are approximated
as particles with energy being less than $\epsilon=2.5 \mbox{ MeV}$ \cite{Zhou:2017zql}.
The chosen cutoff $\epsilon$ is small enough, so that this approximation will not
destroy the initial equilibrium state \cite{Zhou:2017zql}. Initially, the spatial
coordinates of test particles are sampled homogeneously in the box.
While momenta of noncondensate test particles are sampled according to the
Bose-Einstein distribution, energies of condensate test particles are sampled
according to a uniform distribution within the interval $[0,\epsilon]$ and their
momenta are isotropically distributed. 

The time evolution of test particles in phase space is the consequence of 
particle collisions and the free moving between two successive collisions.
In BAMPS, collisions of particles are simulated in a stochastic way according
to collision probabilities corresponding to the collision rates that can be
calculated from the collision term in the Boltzmann equation. Collisions are
realized in each spatial cell with cell length $\Delta L$ within time step $\Delta t$. 
More details on the numerical implementations of collisions $g+g\to g+g$,
$g+g \to g+c$, and $g+c \to g+g$ can be found in Ref. \cite{Zhou:2017zql}.

In our calculations we average $\pi^{xy}({\bf r},t)$ over the space in the box.
Thus, at each time $t$, $\pi^{xy}(t)$ is evaluated as
\begin{equation}
{\pi}^{xy}(t) =\frac{1}{V} \frac{1}{N_{test}} \sum \limits_{i=1}^{N}\frac{p_i^x p_i^y}{E_i}  \,,
\label{eq:tmunu_BAMPS}
\end{equation}
where the sum is over all $N$ noncondensate test particles in the box.
During the time evolution of test particles, $\pi^{xy}$ will only change,
once a collision among test particles occurs, since collisions change
the momenta of colliding particles and thus, weakens the correlation
$\langle \pi^{xy}(t) \pi^{xy}(0) \rangle$. When test particles hit the wall
of the box, they will be moved to the opposite wall, so that they can still
stay in the box. This periodic boundary condition dose not change $\pi^{xy}$,
since the momenta of the test particles do not change.

We calculate $\pi^{xy}(t) \pi^{xy}(0)$ in one run and obtain the correlation
$\langle \pi^{xy}(t) \pi^{xy}(0) \rangle$ by the average over a large number
of runs with different randomly sampled initial conditions. The correlation
will be put into Eq. (\ref{green_kubo_definition}) to calculate the shear
viscosity. For more technical details refer to Ref. \cite{Wesp:2011yy}.

\begin{figure}[ht]
 \centering
 \includegraphics[width=0.45\textwidth]{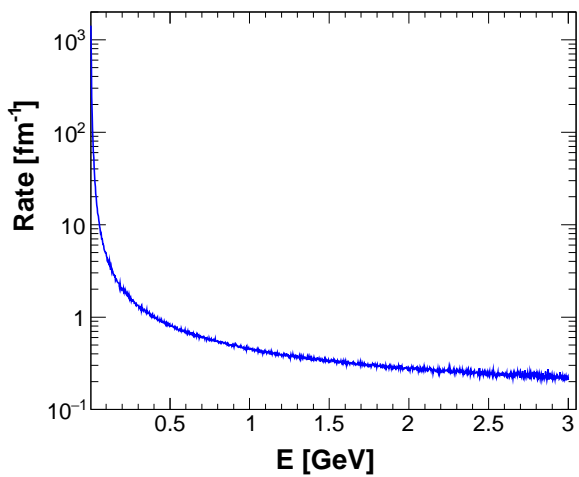}
\caption{Collision rate per real particle as a function of energy.
The results are obtained by employing $\sigma_{22}$ = 1 \mbox{ mb} and 
$N_{test}=2400$. The proportion of BEC accounts for $30\%$.}
\label{fig:prob_p}      
\end{figure}

We present now a new numerical implementation.  As we have noticed
in Sec. \ref{sec2:bec}, the collision rates involving the condensate particle
are infinitely large. With the energy cutoff $\epsilon$ for the condensate particles,
these collision rates are now finite, but still large \cite{Zhou:2017zql}. In addition,
for noncondensate particles, the Bose factor $(1+f_1^g)(1+f_2^g)$ increases
the collision rates significantly, when the momenta of colliding particles are small.
Figure \ref{fig:prob_p} shows the collision rate per particle with energy $E$,
calculated in one time step.
We see that the collision rate increases rapidly, when the energy becomes
small. Therefore, on average, particles with smaller energy have larger
probability to collide than particles with larger energy within one time step.
In BAMPS, we compute the collision probabilities to randomly decide whether
a collision occurs. Since the collision probability is proportional to the time
step times the collision rate and the latter becomes huge for particles with
small energy,  the time step should be chosen quite small, in order to keep
the collision probability smaller than $1$. This leads to a very time
consuming computation, because within one such small time step, particles
with large energy have tiny collision probabilities, but have as much numerical
operations as particles with small energy. Therefore, the numerical handling
for particles with large energy is inefficient. In order to make the computation
more efficient, we introduce in this work two kinds of  time step, a smaller and
a larger one. If the energy of at least one of the two colliding particles is smaller
than a cutoff, $E_{cut}$, the smaller time step is used for calculating the collision
probability. If the energies of both colliding particles are larger than $E_{cut}$,
the larger time step is used. In this way we immensely reduce the computing costs.
In the calculations we choose $E_{cut}=0.5 \mbox{ GeV}$ empirically. 

We list here settings for the numerical calculations. The length of the box
is $L=3 \mbox{ fm}$. The cell length is $\Delta L=0.25 \mbox{ fm}$.
We set $N_{test}=2400$ and perform $200$ independent runs to make
ensemble averages. The temperature of the Boson gas is
$T=0.4 \mbox{ GeV}$. We assume isotropic elastic scatterings. 

\begin{figure}[ht]
 \centering
 \includegraphics[width=0.45\textwidth]{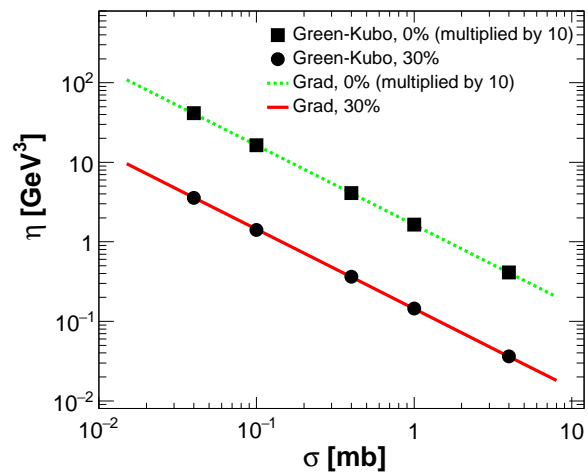}
\caption{Shear viscosity extracted from BAMPS using
the Green-Kubo formalism (symbols). The results from Sec. \ref{sec3:grad}
are shown by the solid and dashed lines.}
\label{fig:eta_cs}      
\end{figure}

Using the Green-Kubo relation, Eq. (\ref{green_kubo_definition}),
the shear viscosity has been calculated for five different cross sections,
as shown in Fig. \ref{fig:eta_cs}. 
The squares depict the results (multiplied by $10$) without a BEC, while
the circles depict the results in the presence of a BEC with
$n_c/(n_c+n_g)=30\%$. The results from the previous section,
Eqs. (\ref{eta_secord}) and (\ref{eq:viscoBEC}), are shown by the solid and 
dashed lines for comparisons. We see excellent agreements between the results 
obtained by using the Green-Kubo relation and the Grad's method of
moments.

We have also calculated the shear viscosity for two further BEC proportions,
$n_c/(n_c+n_g)=15\%, 50\%$. In these cases the total cross section is
set to be $1 \mbox{ mb}$. The results together with those for
$n_c/(n_c+n_g)=0\%, 30\%$ are shown in Fig. \ref{fig:perc} by the circles.
Again, we see perfect agreements between the results from two different
methods.

\section{Shear viscosity of gluons in the presence of a BEC}
\label{sec5:gluon}
For a system of massless gluons, the presence of a gluon BEC would
lower the shear viscosity as expected from the result in the previous sections.
Since the gluon number is not conserved due to number-changing processes
such as two gluons go to three gluons and vice versa, whether a gluon BEC
can be formed (grow) is still under debates 
\cite{Huang:2014iwa,Huang:2013lia,Blaizot:2016iir}.
Taking a first detailed look at the collision terms corresponding to
$g+g \leftrightarrow g +c+c$ processes (which are dominant compared to  
$g+g+g\leftrightarrow g+c$ and $g+g \leftrightarrow g+g+c$), we see that
$f_4^g f_5^g f_1^c f_2^c (1+f_3^g)- f_1^c f_2^c f_3^g (1+f_4^g)(1+f_5^g)$
is zero for massless gluons in equilibrium, while it is negative for massive gluons
with the chemical potential $\mu=m$.
This indicates that number-changing processes cannot destroy a massless
gluon BEC, while they do for a massive gluon BEC.
Following the similar integration that leads to Eq. (\ref{condrate}), we find
that the Bose-Einstein condensation with a finite rate requires an additional
condition: $|{\cal M}_{45\to 123} |^2/s^2$ at $s=4mE-3m^2$ should be nonzero
and finite. For $m=0$ both the rates of $g+g\to g+c+c$ and $g+c+c\to g+g$ are
infinitely large, similar as the rates of $g+g\to g+c$ and $g+c\to g+g$. Therefore,
the same numerical handle with the divergence as done for binary collisions
in the previous section would be used when including these number-changing
processes. On the other hand, for $m >0$, the net rate of the $g+g\to g+c+c$
and $g+c+c\to g+g$ processes is negative, when $\mu=m$. Its value depends on 
$|{\cal M}_{45\to 123} |^2$. Assuming $|{\cal M}_{45\to 123} |^2$ to be constant,
the net rate becomes negative infinite. In this case a BEC may not occur. 
Whether a pQCD motivated $|{\cal M}_{45\to 123} |^2$ will lead to a different
result needs more detailed calculations. This and the effect on the shear
viscosity will be discussed in a future work.
In the current work we assume the dominance of elastic scatterings and ignore
number-changing processes $g+g\leftrightarrow g+c+c$, $g+g+g\leftrightarrow g+c$,
and $g+g\leftrightarrow g+g+c$.
For consistency we also ignore $g+g\leftrightarrow g+g+g$ processes,
although these processes do not destroy the global equilibrium and could be
included to calculate the shear viscosity \cite{Arnold:2003zc,Wesp:2011yy,Uphoff:2014cba}. 
Under these assumptions we expect the (temporarily) presence of a BEC in the early
stages of relativistic heavy-ion collisions and calculate in the following the shear
viscosity of a gluon gas in the presence of a BEC within BAMPS by using the Green-Kubo relation.

We consider a gluon system in thermal equilibrium with a BEC, which
may be formed from a nonequilibrium and overpopulated gluon system
produced as an initial state in ultrarelativistic heavy-ion collisions. 
A simplified form of such an initial distribution of gluons is \cite{Blaizot:2011xf}
\begin{equation}
\label{f_0_init}
f_{init}({\mathbf p})=f_0 \theta (Q_s-|{\mathbf p}|) \,.
\end{equation}
Immediate equilibration would lead to the formation of a BEC for $f_0>0.154$
\cite{Blaizot:2013lga} according to the number and energy conservation.
It was shown in Ref. \cite{Zhou:2017zql} that at equilibrium,
\begin{eqnarray}
\label{nceq}
n_c &=& n_{total} \left [ 1- \zeta(3) \frac{6}{\pi^3} \left ( \frac{15}{4}
\right )^{3/4}  \left ( \frac{1}{f_0} \right )^{1/4} \right ] \,,\\
\label{temp}
T&=& \left ( \frac{15f_0}{4} \right )^{1/4} \frac{Q_s}{\pi} \,,
\end{eqnarray}
where $n_{total}=d_G f_0 Q_s^3/(6 \pi^2)$ is obtained from the initial distribution
(\ref{f_0_init}). $d_G=16$ is the degeneracy factor of gluons.
For $Q_s=1 \mbox{ GeV}$ and $f_0=1.0$ we have $n_c/n_{total}=37\%$
and $T=0.443 \mbox{ GeV}$. If gluon scatterings were isotropic, one could obtain
the effect of the presence of BEC on the gluon shear viscosity according to 
Eqs. (\ref{eta_secord}) and (\ref{eq:viscoBEC}) :
$\eta(37\% \mbox{ BEC})/\eta(\mbox{No BEC})=0.846$.

Different from isotropic scatterings, the matrix element of the elastic
scattering of gluons has the following form:
\begin{equation}
\label{matrix}
| {\cal M}_{gg\to gg} |^2 \approx 144 \pi^2 \alpha_s^2 \frac{s^2}{t (t-m^2_D)} \,,
\end{equation}
which has been calculated by using the Hard-Thermal-Loop (HTL) treatment 
\cite{Aurenche:2002pd,Kurkela:2011ti}.
$s$ and $t$ are the Mandelstam variables and $m_D$ is the Debye screening mass.
In thermal equilibrium we have $m_D=\sqrt{4\pi\alpha_s} T$ \cite{Zhou:2017zql}.
We set $\alpha_s=0.3$. The total cross section is logarithmically divergent, which 
is regularized by an upper cutoff of $t$ \cite{Zhou:2017zql}.

Using BAMPS we evolve an equilibrium gluon system with (or without) a BEC in a box.
We set $T=0.443 \mbox{ GeV}$ and $n_c/n_{total}=37\%, (\mbox{or } 0\%)$. 
The numerical implementations for pQCD scatterings are the same as established 
in Ref. \cite{Zhou:2017zql}. We calculate the shear viscosity of gluons from the Green-Kubo
relation given in the previous section and obtain $\eta/s = 0.438 (\mbox{with BEC})$
and  $\eta/s = 0.531 (\mbox{No BEC})$. Here, $s$ denotes the entropy density.
At equilibrium, $s=d_G 2\pi^2T^3/45$. The gluon BEC has no contribution to the total
entropy. We find that the ratio of the shear viscosity with BEC over the shear viscosity
without BEC is $0.825$, which is almost the same as that calculated with isotropic
scatterings. This suggests that the effect of the presence of BEC on the shear viscosity
is insensitive to the detail of scatterings.

\section{Conclusions}
\label{sec6:sum}
In this paper we have calculated for the first time the shear viscosity of ultrarelativistic
Boson systems in the presence of a BEC. For a special case of massless bosons
with isotropic elastic collisions, the shear viscosity can be derived analytically
by applying the Grad's method, see Eqs. (\ref{eta_secord}) and (\ref{eq:viscoBEC}).
We found that the presence of BEC or more precisely, the interactions
corresponding to the condensation lower the shear viscosity. The larger the
proportion of BEC in the Boson system, the stronger is the reduction of
the shear viscosity. This analytical result is confirmed
by comparing with the numerical result obtained by using the Green-Kubo
relations within the transport approach BAMPS. The agreement between
these results in turn demonstrated the correct numerical implementations
in the BAMPS simulations.

The advantage of the BAMPS simulations over the integrals in the Grad's
method is that the computational expenses in the BAMPS simulations are
same for any form of the matrix element of elastic scatterings, while the
integrals in the Grad's method can be carried out or reduced to integrals
with lower dimensions only for simple forms of the matrix element.
In addition, the current BAMPS simulations can be easily extended to apply
to systems with massive particles.
For a gluon system, the matrix element of elastic scatterings calculated
from pQCD is more complicated than the isotropic form. 
We, thus, use the BAMPS simulations to calculate the shear viscosity of
a gluon system in the presence of a BEC. Our results showed that a
potential formation of a BEC in the early stage of heavy-ion collisions will
reduce the shear viscosity of gluons. The reduction of the shear viscosity 
due to the presence of BEC is insensitive to the detail of interactions.
Thus, the analytical formula of the shear viscosity, Eqs. (\ref{eta_secord}) and 
(\ref{eq:viscoBEC}), obtained  for isotropic scatterings, can be used to 
estimate the effect of the BEC on the shear viscosity of gluons.

\section*{Acknowledgments}
Z.X. thanks D. Rischke for helpful discussions.
This work was financially supported by the National Natural Science
Foundation of China under Grants No. 11575092, No. 11890712, and No. 11335005, 
and the Major State Basic Research Development Program
in China under Grants No.  2014CB845400.  C.G. acknowledges support
by the Deutsche Forschungsgemeinschaft (DFG) through the grant
CRC-TR 211 ``Strong-interaction matter under extreme conditions''.
The BAMPS simulations were performed at Tsinghua National Laboratory
for Information Science and Technology.

\appendix 
\section{Integration in Eq. (\ref{Pmunu})}
\label{app}
In this Appendix we perform the integration $\pi_{\mu\nu} P^{\mu\nu}$ in
Eq. (\ref{Pmunu}) to obtain the result summarized in Eqs. (\ref{eta_secord})
and (\ref{eq:viscoBEC}). 
With $\pi_{\mu\nu}=diag(0, -{\bar \pi}/2, -{\bar \pi}/2, {\bar \pi})$ we have
\begin{eqnarray}
\label{app1}
\pi_{\mu\nu} P^{\mu\nu} &=&2\pi_{\mu\nu} \int d\Gamma_1 p_1^\mu p_1^\nu C[f_i]
\nonumber \\
&=& -{\bar \pi} \int d\Gamma_1 (E_1^2-3p_{1z}^2) C[f_i] \nonumber \\
&=& -\frac{\bar \pi}{4} \int d\Gamma_1 d\Gamma_2 d\Gamma_3 d\Gamma_4 (E_1^2-3p_{1z}^2) 
| {\cal M}_{34\to 12} |^2 \nonumber \\
&& \times \ \left[ f^g_3 f^g_4 (1+f^g_1) (1+f^g_2) 
+f^g_3 f^g_4 (1+f^g_1) f^c_2 \right. \nonumber \\
&& + f^c_3 f^g_4 (1+f^g_1) (1+f^g_2)
+f^g_3 f^c_4 (1+f^g_1) (1+f^g_2) \nonumber \\
&&  - f^g_1 f^g_2 (1+f^g_3) (1+f^g_4) - f^g_1 f^c_2 (1+f^g_3) (1+f^g_4)
\nonumber \\
&&  \left. - f^g_1 f^g_2 f^c_3 (1+f^g_4) - f^g_1 f^g_2 (1+f^g_3) f^c_4 \right] 
\nonumber \\
&& \times (2\pi)^4 \delta^{(4)} (p_3+p_4-p_1-p_2) \,.
\end{eqnarray}
We put $f^g_i=f_i^{BE}[1+(1+f_i^{BE})\phi_i]$ in Eq. (\ref{app1}) with
\begin{equation}
\phi_i=C_0\pi_{i,\mu\nu}p_i^\mu p_i^\nu
=-\frac{\bar \pi}{2} C_0(E_i^2-3p_{iz}^2) \,.
\end{equation}
The sum of integrals with zero power of $\phi_i$ vanishes, since the collision
term vanishes at equilibrium. We calculate integrals with first power of $\phi_i$.
Integrals with higher power are neglected. Therefore, $\pi_{\mu\nu}P^{\mu\nu}$
is proportional to ${\bar \pi}^2$. ${\bar \pi}^2$ from 
$\pi_{\alpha\beta}\pi^{\alpha\beta}$ and $\pi_{\mu\nu}P^{\mu\nu}$ in Eq. (\ref{eta})
cancel out. Thus, $\eta$ does not depend on the magnitude of ${\bar \pi}$.

Some integrals can be carried out analytically. The rest has to be calculated 
numerically. The result is
\begin{eqnarray}
\pi_{\mu\nu} P^{\mu\nu} &=& - \left [ \frac{8}{45}\zeta[5]-\frac{8}{63} \pi^2\zeta[3] 
-\mathcal{A}_1-\mathcal{A}_2 \right. \nonumber \\
&& \left. +\frac{4}{5\pi^2} \frac{n_c}{T^3} (12 \zeta[3]^2+\mathcal{B}) \right ]
{\bar \pi}^2 C_0 \sigma_{22} T^{10} \,,\nonumber \\
&&
\end{eqnarray} 
where 
\begin{eqnarray}
\mathcal{A}_1 &=& \frac{1}{16 T^{10}} \int d\Gamma_3 d\Gamma_4  s^4 
(E_3^2 - 3p_{3z}^2)  (3{\beta^2_z}-1)  \nonumber \\
&& \times f^{BE}_3 f^{BE}_4\int_{-1}^1  du^\prime_1
(1-3u^{\prime 2}_1) (E-P u^\prime_1)^{-4} \nonumber \\
&& \times \left \{  (1+f^{BE}_3) 
f^{BE} \left[ E-s/(E-P u_1^\prime)/2 \right] \right. \nonumber \\
&& +f^{BE}_3 f^{BE} \left[ s/(E-P u_1^\prime)/2 \right] \nonumber \\
&& \left. - f^{BE} \left[ s/(E-P u_1^\prime)/2 \right]^2 \right \} \,,
\end{eqnarray}
\begin{eqnarray}
\mathcal{A}_2 &=& \frac{1}{T^{10}} \int d\Gamma_1 d\Gamma_2 s 
(E_1^2 - 3p_{1z}^2+E_2^2 - 3p_{2z}^2) \nonumber \\
&& \times (E_1^2 - 3p_{1z}^2)
\left \{ \frac{T}{P} \left [ f^{BE}(E)-f^{BE}_2 \right ] \right. \nonumber \\
&& \times \left. \ln \frac{f^{BE}[(E-P)/2]}{f^{BE}[(E+P)/2]}
-\frac{1}{2} f^{BE}(E) \right \}
\nonumber \\ 
&& \times f^{BE}_1(1+f^{BE}_1) \,,
\end{eqnarray}
\begin{equation}
\mathcal{B} = \int_0^1 dx_1 \int_0^1 dx_2  
\frac{x_1(\ln x_1)^3}{(1-x_1)^2}\frac{(\ln x_2)^2}{1-x_1x_2} \,. 
\end{equation}
Remember that $E=E_3+E_4$ is the total energy and
${\bf P}={\bf p}_3+{\bf p}_4$ is the total momentum. 
$\mathcal{A}_1$, $\mathcal{A}_2$, and $\mathcal{B}$ can only be
calculated numerically. 
In the following we carry out explicitly three integrals in Eq. (\ref{app1})
as examples.

\subsection{First integral}
\begin{eqnarray}
\mathcal{I}_1 &=& -\frac{\bar \pi}{4} \int d\Gamma_1 d\Gamma_2 d\Gamma_3 d\Gamma_4 
( E_1^2 - 3 p_{1z}^2 ) | {\cal M}_{34\to 12} |^2 \nonumber \\
&& \times 2f^{BE}_3f^{BE}_4(1+f^{BE}_3)\phi_3
(2\pi)^4 \delta^{(4)} (p_3+p_4-p_1-p_2) \nonumber \\ 
&=& \frac{1}{4} {\bar \pi}^2 C_0  \int d\Gamma_1 d\Gamma_2 d\Gamma_3 d\Gamma_4 
( E_1^2 - 3 p_{1z}^2 ) (E_3^2-3p_{3z}^2) \nonumber \\ 
&& \times | {\cal M}_{34\to 12} |^2 f^{BE}_3f^{BE}_4(1+f^{BE}_3) \nonumber \\
&& \times (2\pi)^4 \delta^{(4)} (p_3+p_4-p_1-p_2) \,.
\label{I1_1}
\end{eqnarray}
This integral corresponds to scatterings among noncondensate particles, $g+g \to g+g$.
Remember that $| {\cal M}_{34\to 12} |^2=32\pi s\sigma_{22}$ for isotropic elastic
scatterings, where $s=(p_1+p_2)^2=(p_3+p_4)^2$. We have assumed constant cross section
$\sigma_{22}$.

At first, we integrate over $d\Gamma_2$ with help of 
$\delta^{(3)} ({\bf p}_3+{\bf p}_4-{\bf p}_1-{\bf p}_2)$ and obtain
\begin{eqnarray}
\mathcal{I}_1 &=& \frac{\pi}{2} {\bar \pi}^2 C_0 \int d\Gamma_3 d\Gamma_4
(E_3^2-3p_{3z}^2) | {\cal M}_{34\to 12} |^2 f^{BE}_3f^{BE}_4 \nonumber \\
&& \times (1+f^{BE}_3) \int d\Gamma_1 \frac{E_1^2 - 3p_{1z}^2}{2(E-E_1)}
\delta[F({\bf p}_1)] \,.
\label{I1_2}
\end{eqnarray}
$F({\bf p}_1)$ indicates the energy conservation
\begin{equation}
F({\bf p}_1)=E-E_1-E_2=E-E_1-\sqrt{({\bf P}-{\bf p}_1)^2} \,.
\label{I1_3}
\end{equation}

Second we integrate over $d\Gamma_1$. To do it, we rotate the coordinate
system ${\bf \hat p}$ to a new one ${\bf \hat p^\prime}$ with ${\bf \hat p^\prime}_z$
paralleling to ${\bf P}$. In the new coordinate system, $E$, $|{\bf P}|\equiv P$,
$E_1$, and the angle between ${\bf P}$ and ${\bf p}_1$ are unchanged.
$p_{1z}$ is transferred to 
\begin{equation}
 p_{1z}=-p^\prime_{1x}\beta_x-p^\prime_{1y}\beta_y+p^\prime_{1z}\beta_z
\label{I1_4}
\end{equation}  
where $\beta_x$, $\beta_y$, and $\beta_z$ are the cosine of angles between $\bf P$
and the old coordinate axes, ${\bf \hat p}_x$, ${\bf \hat p}_y$, and ${\bf \hat p}_z$,
respectively. It is easy to prove that $\beta_x^2+\beta_y^2+\beta_z^2=1$.
The integral over $d\Gamma_1$ in Eq. (\ref{I1_2}) is evaluated in the new
coordinate system 
\begin{eqnarray}
&&\int d\Gamma_1^\prime \frac{1}{2(E-E_1^\prime)}
[{E_1^\prime}^2 - 3(-p^\prime_{1x}\beta_x-p^\prime_{1y}\beta_y+p^\prime_{1z}\beta_z)^2]
\nonumber \\
&& \times \delta[F({\bf p}_1^\prime)] \nonumber \\
&=& \int \frac{{p_1^\prime}^2 dp^{\prime}_1 \sin{\theta}^\prime_1 d{\theta}^\prime_1 
d{\phi}^\prime_1} {(2\pi)^3 2E^{\prime}_1}\frac{1}{2(E-E^{\prime}_1)} {E_1^\prime}^2 [1 \nonumber \\
&& - 3{(\sin{\theta}^\prime_1 \cos{\phi}^\prime_1 \beta_x+\sin{\theta}^\prime_1 
\sin{\phi}^\prime_1 \beta_y -\cos{\theta}^\prime_1 \beta_z)}^2 ] \nonumber \\
&&\times \delta[F({\bf p}_1^\prime)] \,,
\label{I1_5}
\end{eqnarray}
where
\begin{equation}
F({\bf p}_1^\prime)=E-E_1^\prime-\sqrt{P^2+{E_1^\prime}^2-2PE_1^\prime \cos \theta_1^\prime} \,.
\label{I1_6}
\end{equation}
By integrating over $\phi_1^\prime$, Eq. (\ref{I1_5}) is equal to
\begin{equation}
\frac{3\beta_z^2-1}{32\pi^2} \int dE^{\prime}_1 d\cos{\theta}^\prime_1 
\frac{{E_1^\prime}^3}{E-E^{\prime}_1} (1-3\cos^2\theta_1^\prime)
\delta[F({\bf p}_1^\prime)] \,.
\label{I1_7}
\end{equation}
With
\begin{equation}
\delta[F]=\frac{\delta(E^\prime_1-{\bar E^\prime_1})}
{\vert \partial F/\partial E^\prime_1 \vert} 
=\frac{E-E^\prime_1}{E-P \cos{\theta}^\prime_1}
\delta(E^\prime_1-{\bar E^\prime_1})\,,
\label{I1_8}
\end{equation}
where ${\bar E^\prime_1}=(E^2-P^2)/[2(E-P \cos{\theta}^\prime_1)]$ is
the solution of $E^\prime_1$ in $F=0$, we perform the integral
in Eq. (\ref{I1_7}) over $E_1^\prime$ and obtain
\begin{eqnarray}
&&\frac{3\beta_z^2-1}{256\pi^2}  (E^2-P^2)^3 \int_{-1}^1 d \cos\theta^{\prime}_1
\frac{1-3\cos^2\theta_1^\prime}{(E-P\cos\theta_1^\prime)^4} 
\nonumber \\
&=& \frac{1-3\beta_z^2}{48\pi^2} P^2 \,.
\label{I1_9}
\end{eqnarray}

Putting the integral over $\Gamma_1$, Eq.  (\ref{I1_9}), and
$| {\cal M}_{34\to 12} |^2=32\pi s\sigma_{22}$ into Eq. (\ref{I1_2}),
we have 
\begin{eqnarray}
\mathcal{I}_1 &=& \frac{1}{3} {\bar \pi}^2 C_0 \sigma_{22} 
\int \frac{d^3 p_3}{(2\pi)^3 2E_3} \frac{d^3 p_4}{(2\pi)^3 2E_4} 
f^{BE}_3f^{BE}_4  (1+f^{BE}_3) \nonumber \\
&& \times (E_3^2-3p_{3z}^2) s (1-3\beta_z^2) P^2 \,,
\label{I1_10}
\end{eqnarray}
where $s=E^2-P^2$, 
$\beta_z={\bf P} \cdot {\bf \hat p}_z/P=(p_{3z}+p_{4z})/P$,
and 
\begin{eqnarray}
\label{I1_11}
P&=&({\bf p}_3+{\bf p}_4)^{1/2}= \left \{ p_3^2+p_4^2+
2p_3p_4[\cos\theta_3 \cos\theta_4 \right. \nonumber \\
&&\left. + \sin\theta_3\sin\theta_4\cos(\phi_3-\phi_4)] \right \}^{1/2} \,.
\end{eqnarray}
Integrating over $\phi_3$ and $\phi_4$ gives
\begin{eqnarray}
\mathcal{I}_1 &=& \frac{1}{96\pi^4} {\bar \pi}^2 C_0 \sigma_{22} 
\int dE_3 du_3 dE_4 du_4 f^{BE}_3f^{BE}_4  (1+f^{BE}_3) \nonumber \\
&& \times E_3^4 E_4^2 (1-3u_3^2) [3E_3E_4u_3^2u_4^2-
(3E_3^2-E_3E_4)u_3^2 \nonumber \\
&& -(3E_4^2-E_3E_4)u_4^2+(E_3^2+E_4^2+E_3E_4) ]\,,
\label{I1_12}
\end{eqnarray}
where $u_3=\cos\theta_3$ and $u_4=\cos\theta_4$. By further integral
over $u_3$, $u_4$, $E_3$, and $E_4$ we obtain
\begin{eqnarray}
\mathcal{I}_1 &=& \frac{1}{96\pi^4} {\bar \pi}^2 C_0 \sigma_{22} 
\int dE_3 dE_4 f^{BE}_3f^{BE}_4  (1+f^{BE}_3) \nonumber \\
&& \times E_3^4 E_4^2 \left ( \frac{16}{5} E_3^2-\frac{32}{15}E_3E_4
\right ) \nonumber \\
&=& \frac{8}{315} (2\pi^2\zeta[3]-7\zeta[5])  {\bar \pi}^2 C_0 \sigma_{22} T^{10} \,.
\label{I1_13}
\end{eqnarray}

\subsection{Second integral}
\begin{eqnarray}
\mathcal{I}_2 &=& -\frac{\bar \pi}{4} \int d\Gamma_1 d\Gamma_2 d\Gamma_3 d\Gamma_4 
( E_1^2 - 3 p_{1z}^2 ) | {\cal M}_{34\to 12} |^2 2f^{BE}_2 \nonumber \\
&& \times f^{BE}_3f^{BE}_4(1+f^{BE}_3)\phi_3
(2\pi)^4 \delta^{(4)} (p_3+p_4-p_1-p_2) \nonumber \\ 
&=& \frac{1}{4} {\bar \pi}^2 C_0  \int d\Gamma_1 d\Gamma_2 d\Gamma_3 d\Gamma_4 
( E_1^2 - 3 p_{1z}^2 ) (E_3^2-3p_{3z}^2) \nonumber \\ 
&& \times | {\cal M}_{34\to 12} |^2 f^{BE}_2 f^{BE}_3f^{BE}_4(1+f^{BE}_3) \nonumber \\
&& \times (2\pi)^4 \delta^{(4)} (p_3+p_4-p_1-p_2) \,.
\label{I2_1}
\end{eqnarray}
The only difference of $\mathcal{I}_2$ from $\mathcal{I}_1$ is the additional
multiplier $f^{BE}_2(E_2)$, which changed to $f^{BE}(E-E_1)$ by the integral
over $\Gamma_2$ due to the energy conservation. We follow the integration in
the previous section until Eq. (\ref{I1_9}). Instead of Eq. (\ref{I1_9})
we have now
\begin{eqnarray}
&& \frac{3\beta_z^2-1}{256\pi^2}  (E^2-P^2)^3 \int_{-1}^1 d \cos\theta^{\prime}_1
\frac{1-3\cos^2\theta_1^\prime}{(E-P\cos\theta_1^\prime)^4} \nonumber \\ 
&& \times f^{BE}(E-{\bar E}_1^\prime) \nonumber \\
&=&\frac{3\beta_z^2-1}{256\pi^2}  (E^2-P^2)^3 \int_{-1}^1 d \cos\theta^{\prime}_1
\frac{1-3\cos^2\theta_1^\prime}{(E-P\cos\theta_1^\prime)^4} \nonumber \\ 
&& \times f^{BE}[E-s/(E-Pu_1^\prime)/2]
\label{I2_2}
\end{eqnarray}
which can only be calculated numerically.

\subsection{Third integral}
\begin{eqnarray}
\mathcal{I}_3 &=& -\frac{\bar \pi}{4} \int d\Gamma_1 d\Gamma_2 d\Gamma_3 d\Gamma_4 
( E_1^2 - 3 p_{1z}^2 ) | {\cal M}_{34\to 12} |^2 2 f^c_2 \nonumber \\
&& \times f^{BE}_3f^{BE}_4(1+f^{BE}_3)\phi_3
(2\pi)^4 \delta^{(4)} (p_3+p_4-p_1-p_2) \nonumber \\ 
&=& 2\pi^3 {\bar \pi}^2 C_0 n_c \int d\Gamma_1 d\Gamma_2 d\Gamma_3 d\Gamma_4 
( E_1^2 - 3 p_{1z}^2 ) \nonumber \\ 
&& \times  (E_3^2-3p_{3z}^2)  | {\cal M}_{34\to 12} |^2 \delta^{(3)}({\bf p}_2) f^{BE}_3f^{BE}_4(1+f^{BE}_3)
\nonumber \\
&& \times (2\pi)^4 \delta^{(4)} (p_3+p_4-p_1-p_2) \,.
\label{I3_1}
\end{eqnarray}
The only difference of $\mathcal{I}_3$ from $\mathcal{I}_1$ is the additional
multiplier $f^c_2=(2\pi)^3 n_c \delta^{(3)}({\bf p}_2)$. Thus, this integral
corresponds to scatterings between condensate and noncondensate particles,
$g+g\to c+g$ or $c+g\to g+g$. Remember that for these scatterings to occur,
$| {\cal M}_{34\to 12} |^2 /s$ should be finite at $s=2mE$, 
see Eq. (\ref{condrate}). For isotropic scatterings 
$| {\cal M}_{34\to 12} |^2 /s=32\pi\sigma_{22}$ is finite for finite
cross sections.

We integrate first over $d\Gamma_1$ with help of 
$\delta^{(3)} ({\bf p}_3+{\bf p}_4-{\bf p}_1-{\bf p}_2)$ and obtain
\begin{eqnarray}
\mathcal{I}_3 &&= 4\pi^4 {\bar \pi}^2 C_0 n_c \frac{| {\cal M}_{34\to 12} |^2}{s}
\int d\Gamma_3 d\Gamma_4 (E_3^2-3p_{3z}^2) 2mE  \nonumber \\
&& \times f^{BE}_4 f^{BE}_3(1+f^{BE}_3) \int d\Gamma_2 
\frac{(E-E_2)^2 - 3(P_z-p_{2z})^2}{2(E-E_2)} \nonumber \\
&& \times  \delta^{(3)}({\bf p}_2) \delta[F({\bf p}_2)]\,,
\label{I3_2}
\end{eqnarray}
where
\begin{equation}
F({\bf p}_2)=E-E_2-\sqrt{({\bf P}-{\bf p}_2)^2+m^2}\,.
\label{I3_3}
\end{equation}
$m$ is the rest mass of particles. We will let $m$ to equal zero at the end
of the integration.

Using the identity
\begin{equation}
\int dE_2d^3p_2 \delta (E^2_2-p^2_2-m^2)= \int \frac{d^3 p_2}{2E_2} \,,  
\label{I3_4}
\end{equation} 
the integral over $\Gamma_2$ is expressed to
\begin{eqnarray}
&& \int \frac{1}{(2\pi)^3} dE_2 d^3 p_2
\frac{(E-E_2)^2 - 3(P_z-p_{2z})^2}{2(E-E_2)}\nonumber \\
&& \times \delta(E_2^2-p_2^2-m^2)\delta^{(3)}({\bf p}_2)\delta[F({\bf p}_2)]\,.
\label{I3_5}
\end{eqnarray}
The integration over ${\bf p}_2$ gives
\begin{eqnarray}
&& \int \frac{1}{(2\pi)^3} dE_2 \frac{(E-E_2)^2 - 3P_z^2}{2(E-E_2)}
\delta(E_2^2-m^2) \nonumber \\
&& \times \delta(E-E_2-\sqrt{P^2+m^2}) \nonumber \\
&=& \frac{1}{(2\pi)^3} \frac{P^2 +m^2- 3P_z^2}{2\sqrt{P^2+m^2}}
\delta[(E-\sqrt{P^2+m^2})^2-m^2] \,. \nonumber \\
\label{I3_6}
\end{eqnarray}

Putting the integral over $\Gamma_2$, Eq.  (\ref{I3_6}), into Eq. (\ref{I3_2}),
we have 
\begin{eqnarray}
\mathcal{I}_3 &=& 8\pi^2 {\bar \pi}^2 C_0 \sigma_{22} n_c \int d\Gamma_3 d\Gamma_4
 f^{BE}_4  f^{BE}_3(1+f^{BE}_3) \nonumber \\
&& \times (E_3^2-3p_{3z}^2) 2m E \frac{P^2 +m^2- 3P_z^2}{\sqrt{P^2+m^2}} \nonumber \\
&& \times \delta[(E-\sqrt{P^2+m^2})^2-m^2] \,.
\label{I3_7}
\end{eqnarray}
We define $G=(E-\sqrt{P^2+m^2})^2-m^2$.  According to Eq. (\ref{I1_11}) we obtain 
\begin{eqnarray}
\label{I3_8}
\delta(G)&=&\sum^{2}_{k=1} \frac{\delta (\phi_4- {\bar \phi}_{4,k})}{| \partial G/ \partial \phi_4|}
\nonumber \\
&=&\frac{\sqrt{P^2+m^2}\sum^{2}_{k=1} \delta (\phi_4-{\bar \phi}_{4,k})}
{2m p_3p_4 \sqrt{1-A^2+2Au_3u_4-u^2_3-u^2_4}} \,, \nonumber \\
&&
\end{eqnarray}
where $A\equiv (E_3-m)(E_4-m)/p_3/p_4$ and ${\bar \phi}_{4,1}$ and ${\bar \phi}_{4,2}$
are two solutions of $\phi_4$ in $G=0$, since cosine of $\alpha_1$ and
$\alpha_2=2\pi-\alpha_1$ are equal. $G=0$ also leads to $P^2+m^2=(E-m)^2$.
The integral over $\phi_4$ gives a factor of $2$ because of two solutions and the
further integral over $\phi_3$ gives a factor of $2\pi$. Then we have
\begin{eqnarray}
\mathcal{I}_3 &=& \frac{1}{(2\pi)^3} {\bar \pi}^2 C_0 \sigma_{22} n_c 
\int dp_3 dp_4 \frac{p_3p_4}{E_3E_4} E f^{BE}_4  f^{BE}_3 \nonumber \\
&& \times (1+f^{BE}_3) \int du_3 du_4 (E_3^2-3p_3^2u_3^2)  \nonumber \\
&& \times \frac{(E-m)^2-3(p_3u_3+p_4u_4)^2}
{\sqrt{1-A^2+2Au_3u_4-u^2_3-u^2_4}} \,.
\label{I3_9}
\end{eqnarray}

The upper and lower boundary of $u_4$ are the solutions of 
$1-A^2+2Au_3u_4-u^2_3-u^2_4=0$. Beyond these boundaries the square root 
in Eq. (\ref{I3_9}) is negative. We integrate over $u_4$ and obtain
\begin{eqnarray}
\mathcal{I}_3 &=& \frac{1}{(2\pi)^3} {\bar \pi}^2 C_0 \sigma_{22} n_c 
\int dp_3 dp_4 \frac{p_3p_4}{E_3E_4} E f^{BE}_4  f^{BE}_3 \nonumber \\
&& \times (1+f^{BE}_3) \int du_3  (E_3^2-3p_3^2u_3^2)  
 \pi \left \{ (E-m)^2 \right. \nonumber \\
&& \left.  - 3u^2_3p^2_3-\frac{3}{2}[1-A-(1-3A^2)u_3^2]p^2_4 -6Au^2_3p_3p_4 \right \}
\,. \nonumber \\
&&
\label{I3_10}
\end{eqnarray}
We now let the mass $m$ to be zero, which leads to $A=1$, $p_3=E_3$,
$p_4=E_4$. With these we perform the integration in Eq. (\ref{I3_10}) over
$u_3$, $E_3$, and $E_4$ and obtain  
\begin{eqnarray}
\mathcal{I}_3 &=& \frac{1}{8\pi^2} {\bar \pi}^2 C_0 \sigma_{22} n_c 
\int dE_3 dE_4 (E_3+E_4)^3E_3^2 f^{BE}_4  f^{BE}_3   \nonumber \\
&& \times (1+f^{BE}_3) \int_{-1}^1 du_3  (1-3u_3^2)^2  \nonumber \\
&=& \frac{1}{5\pi^2} {\bar \pi}^2 C_0 \sigma_{22} n_c 
\int dE_3 dE_4 (E_3+E_4)^3E_3^2 f^{BE}_4  f^{BE}_3 \nonumber \\
&& \times  (1+f^{BE}_3) \nonumber \\
&=&\frac{1}{5\pi^2} \left ( \frac{7\pi^6}{45}+36\zeta[3]^2 
+5! \zeta[5] \int_0^\infty dx \frac{1}{e^x-1} \right ) \nonumber \\
&& \times {\bar \pi}^2 C_0 \sigma_{22} n_c T^7 \,.
\label{I3_11}
\end{eqnarray}
The integral over $x$ is logarithmically divergent, which
will cancel with the corresponding term of the integral
\begin{eqnarray}
&& \frac{\bar \pi}{4} \int d\Gamma_1 d\Gamma_2 d\Gamma_3 d\Gamma_4 
( E_1^2 - 3 p_{1z}^2 ) | {\cal M}_{34\to 12} |^2 2 f^c_4 \nonumber \\
&& \times f^{BE}_1f^{BE}_2(1+f^{BE}_1)\phi_1
(2\pi)^4 \delta^{(4)} (p_3+p_4-p_1-p_2)  \,. \nonumber \\
&&
\end{eqnarray}

\end{document}